\documentclass[sigconf]{acmart}
\renewcommand{\theenumii}{\theenumi.\arabic{enumii}.}

\renewcommand\footnotetextcopyrightpermission[1]{}  

\setcopyright{none}
\acmDOI{}

\usepackage[table,xcdraw]{xcolor}
\usepackage{multirow}
 \usepackage{graphicx}
\usepackage{multirow}
\usepackage{amsmath}
\usepackage{enumitem}
\usepackage{array}
\usepackage{lscape}
\usepackage{booktabs}
\newcolumntype{P}[1]{>{\centering\arraybackslash}p{#1}}
\usepackage{longtable}
\usepackage[table]{xcolor}
\usepackage{tikz}
\usepackage{array}
\usepackage{soul}

\newcommand{\rulesep}{\unskip\ \vrule\ }

\newcommand{\multirowvertlinecustom}[4]{%
  \multirow{#1}{*}{%
    \begin{tikzpicture}[baseline=(char.base)]
      \node[inner sep=0pt] (char) {\strut};
      \draw[line width=0.3mm] (#4,#2) -- (#4,#3);
    \end{tikzpicture}%
  }%
}

\begin{document}
\title{Graph Neural Team Recommendation: An Integrated Approach}

\author{Md Jamil Ahmed}
\orcid{0009-0005-3185-4046}
\affiliation{%
  \institution{University of Windsor}
  \city{Windsor}
   \state{Ontario}
 \country{Canada}
 }
 \email{ahmed491@uwindsor.ca} 

\author{Mahdis Saeedi}
\orcid{0000-0002-6297-3794}
\affiliation{%
  \institution{University of Windsor}
 \city{Windsor}
 \state{Ontario}
 \country{Canada}
}
\email{msaeedi@uwindsor.ca}

\author{Hossein Fani}
\orcid{0000-0002-6033-6564}
\affiliation{%
  \institution{University of Toronto}
 \city{Toronto}
 \state{Ontario}
 \country{Canada}
}
\email{hossein.fani@utoronto.ca}


\begin{abstract}
Team recommendation aims to select an optimal subset of experts who can form an \textit{almost surely} successful collaborative team for a given set of required skills. State-of-the-art methods are neural multi-label classifiers that \textit{transfer} dense vector representations of skills into a sparse occurrence vector representing the optimal subset of experts. Such methods, however, overlook experts' relational and structural information encoded in the expert collaboration graph and, thus, fall short of capturing complex inter-dependencies among experts and their associated skills within teams. Moreover, the skills' dense vectors are pretrained \textit{disjointly} and independently of the underlying neural classifier, hence, preventing end-to-end optimization. In this paper, we propose to reformulate the team recommendation problem into \textit{end-to-end} link predictions in the expert collaboration graph to consume multi-hop intra-team and cross-team collaborations among experts while eschewing the unnecessary complexities of the disjoint two-phase training procedure. Our experiments on two large-scale datasets from various domains with distinct distributions of skills in teams demonstrate the superiority of the end-to-end approach and establish a new state of the art. Our code is available at \texttt{https://github.com/fani-lab/OpeNTF}.
\vspace{-0.5em}
\end{abstract}
\vspace{-1em}
\keywords{Neural Team Recommendation; Social Information Retrieval;}
\vspace{-1em}

\ccsdesc[300]{Human-centered computing~Social recommendation\vspace{-0.85em}}


\pagestyle{plain} 
\maketitle
\begin{figure}[t]
\centering
\includegraphics[width=0.37\columnwidth]{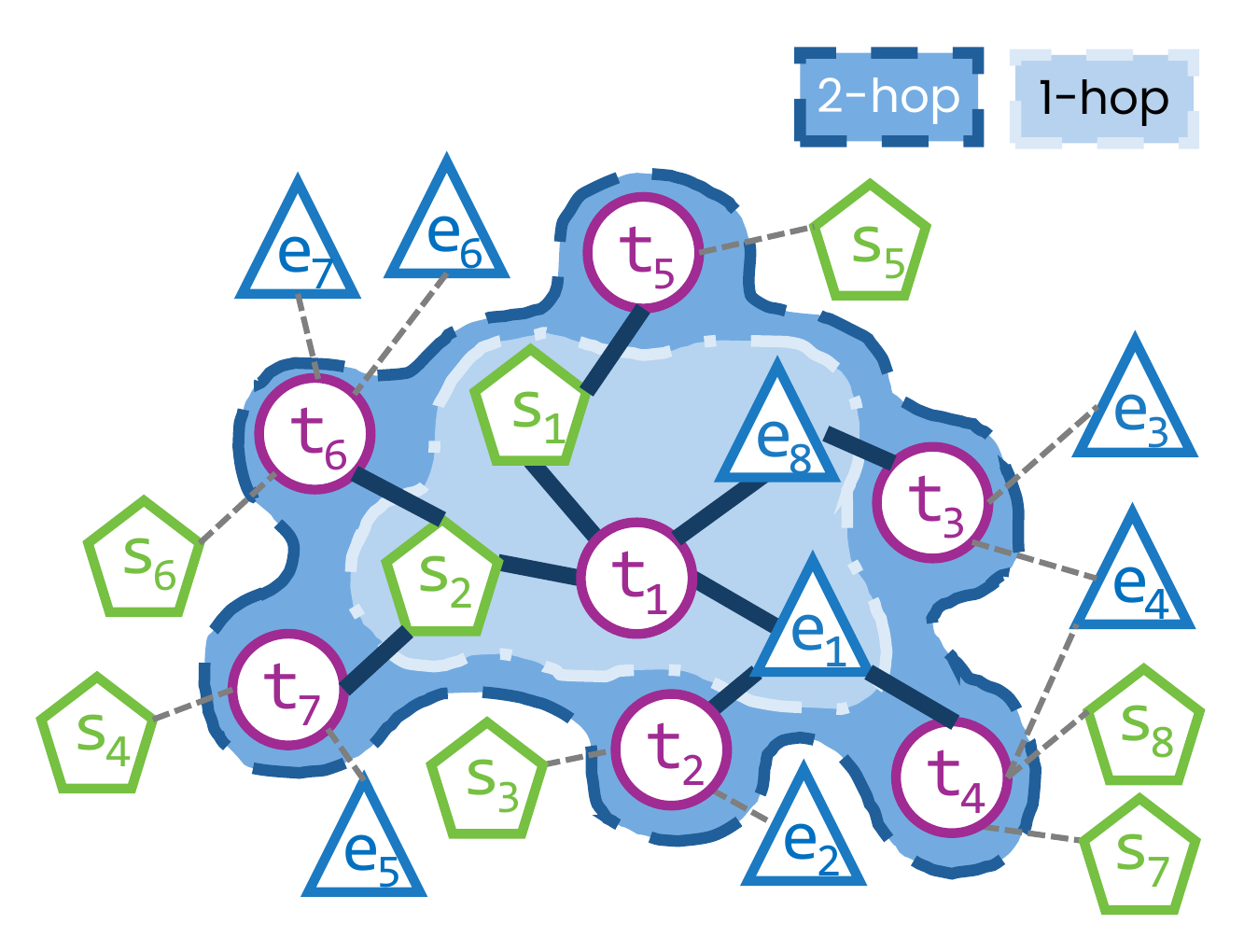}
\vspace{-1.5em}
\vrule width 0.5pt
\includegraphics[width=0.61\columnwidth]{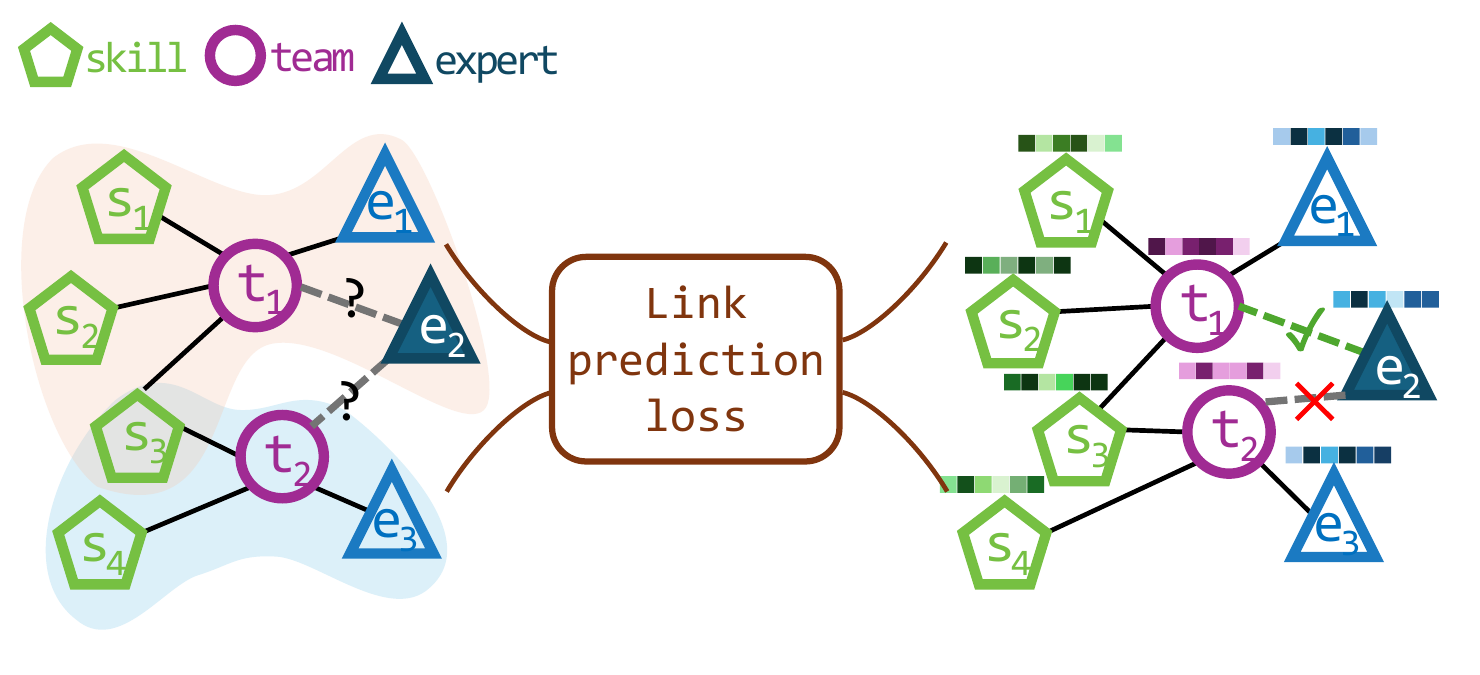}
\caption{Multi-hop relations and structure in the expert collaboration graph (left). Our end-to-end approach (right).}
\label{fig:hop-e2e}
\vspace{-2.6em}
\end{figure}

\vspace{-1.5em}
\section{Introduction}
Collaborative teamwork has become paramount in diverse real-world settings, including health care~\cite{DBLP:conf/inista/SelvarajahZKKIP18,mcleod2021hacking}, scientific peer review~\cite{DBLP:journals/jair/AksoyYA23,DBLP:conf/cikm/ArabzadehESBB24,DBLP:journals/rpjdi/KreutzS23}, and education~\cite{michaelsen2023team,DBLP:conf/hais/CandelSAJB23}, where combined skills applied in coordinated ways can solve difficult tasks. As a result, forming teams of experts whose success is \textit{almost surely} guaranteed has been a surge of research for years; from operations research~\cite{DBLP:journals/asc/StrnadG10,DBLP:journals/eswa/WiOMJ09,ZakarianK99,DBLP:journals/anor/CampeloFS20,DBLP:journals/fgcs/DurfeeBS14,KalayathankalAVK19,DBLP:journals/access/WangZCPC20,DBLP:journals/corr/abs-1903-03523}, social network analysis~\cite{DBLP:conf/cikm/KargarA11,DBLP:conf/kdd/SozioG10,DBLP:conf/icde/GolzadehGS25}, and more recently, machine learning~\cite{DBLP:conf/ecir/FaniBDS24, DBLP:conf/cikm/DashtiSF22, DBLP:conf/cikm/RadFKSB20,DBLP:conf/sigir/RadBKSS21}. Specifically, neural models~\cite{DBLP:conf/cikm/RadMFKSB21,DBLP:conf/sigir/RadBKSS21,DBLP:journals/ir/RadNABKSSZ23,DBLP:conf/cikm/DashtiSPF22,DBLP:conf/cikm/DashtiSF22,DBLP:journals/kbs/RueninCST24} have brought state-of-the-art efficacy while enhancing efficiency due to the iterative learning procedure and availability of training datasets. Neural models, by and large, frame the team recommendation problem as a \textit{multi-label} Boolean classification task and learn the vector representations of experts and their skills in a training dataset of successful and (virtually) unsuccessful teams to draw future teams that are, more likely than not, successful. Such work, however, was premised on the mutually independent selection of teams from a training set, overlooking the fact that a team is indeed inherently relational, and its success depends on the intra- and cross-team collaborations among experts. 

In this paper, we propose to model collaborative ties of experts within and across teams in a graph and leverage graph neural networks to capture multi-hop relational and structural information encoded in the graph for complex inter-dependencies among experts and their associated skills within teams, as shown in Figure~\ref{fig:hop-e2e}. We then propose to reformulate the team recommendation problem into end-to-end link predictions between nodes of required skills and optimal experts for a team node directly within the graph. Given a successful team, we map it into a connected \textit{star} subgraph that connects the team's node to its required subset of skills and expert members. After incorporating all successful teams of a training set in the graph, we then apply a graph neural network to learn dense vectors of skills and experts jointly based on link prediction loss among positive (existing) and negative (non-existing) link sampling strategies. During inference, given a test team with its subset of required skills yet \textit{unseen} expert members, as shown in Figure~\ref{fig:hop-e2e} (right), we use the graph neural network to predict links between expert nodes and the team's node and select the top-$k$ highest probable experts as the recommended team of size $k$.

While employing graphs to model experts' collaborations was mainstream in traditional team recommendation approaches rooted in subgraph optimization~\cite{DBLP:conf/cikm/KargarA11,DBLP:conf/kdd/SozioG10,DBLP:conf/icde/GolzadehGS25}, existing neural-based methods have leveraged graphs only for prior pretraining and the \textit{transfer} of dense vectors of skills to accommodate a large set of skills and reduce the complexity of the neural classifiers in the input layer~\cite{DBLP:journals/ir/RadNABKSSZ23, DBLP:conf/sigir/RadBKSS21, DBLP:conf/ijcnn/KawKS23}. 
Such approaches, however, entail key shortcomings: (1) foremost, they overlook the multi-hop relational and structural information among experts and their collaborations within and across teams in the graph; (2) the skill vectors are learned \textit{disjointly} in a self-supervised manner during a pretraining phase, oblivious to the team recommendation learning process; (3) they suffer from independent selection of teams from the training set by the underlying neural classifiers. 

Inspired by the efficacy of end-to-end graph neural networks in various recommendation and information retrieval tasks~\cite{DBLP:conf/kdd/WangLLYW20,DBLP:conf/aaai/HuangXXDXLBXLY21,DBLP:conf/www/Fan0LHZTY19,DBLP:conf/www/ChenBSXZHHWH24,DBLP:journals/tors/GaoZLLQPQCJHL23,DBLP:conf/nips/ZhuZXT21,DBLP:conf/sigir/0004Z0LSWWCKA24,DBLP:conf/naacl/LiGSSBZW25,DBLP:conf/naacl/HuLZPLZ25,DBLP:conf/cikm/KimCYK21}, yet unexplored in team recommendation, our approach fills the gaps by capturing multi-hop relational and structural information not only for skills but also for experts and teams directly in the expert collaboration graph, thereby recommending more effective teams as evidenced by our experiments, including \texttt{3} different graph structures, \texttt{6} strong graph neural networks, and \texttt{2} large-scale datasets from diverse domains with varied distributions of teams over skills.

\vspace{-0.5em}
\section{Problem Definition}\label{sec:problem_definition}
Given a set of skills $ \mathcal{S} = \{s_i \} $ and a set of experts $ \mathcal{E} = \{e_j\} $, a team $ t $ is a subset of experts $ \textbf{e} \subseteq \mathcal{E} $ that collectively cover a subset of skills $ \textbf{s} \subseteq \mathcal{S}$ with its success status $y\in\{0,1\}$, denoted by $t_{\textbf{s},\textbf{e},y} $. Further, $\mathcal{T}=\mathcal{T}^+\cup \mathcal{T}^- =\{t_{\textbf{s},\textbf{e},y=1}\}\cup\{t_{\textbf{s},\textbf{e},y=0}\}$ is the set of successful and unsuccessful teams. For a subset of skills $\textbf{s}$, a neural team recommender aims to identify an optimal subset of experts $\textbf{e}$ such that their collaboration results in a successful team, i.e., $t_{\textbf{s},\textbf{e},y=1}$, while avoiding any subset $\textbf{e}'$ that leads to an unsuccessful team, i.e., $t_{\textbf{s},\textbf{e}',y=0}$. Specifically, the objective is to learn a mapping function $f$, parameterized by $\theta$, such that $\forall t\in \mathcal{T}^+$: $f_{\theta}(\textbf{s}) = \textbf{e}$.


\vspace{-0.5em}
\section{Preliminaries} \label{sec:ste}
The state of the art~\cite{DBLP:conf/cikm/RadFKSB20,DBLP:journals/ir/RadNABKSSZ23,DBLP:conf/sigir/RadBKSS21,DBLP:conf/ijcnn/KawKS23,DBLP:conf/cikm/DashtiSF22} estimate $f_\theta(\textbf{s})=\textbf{e}$ using a neural multi-label Boolean classifier that transfers the \textit{dense} vector representation of a subset of skills, $v_\textbf{s}\in \mathbb{R}^d$, to the occurrence (multi-hot) vector representation of the optimal (successful) subset of experts, $v_\textbf{e}\in \{0,1\}^{|\mathcal{E}|}$, i.e., $f_\theta(v_\textbf{s}) = v_\textbf{e}$, by maximizing the posterior probability of $\theta$ in $ f_\theta$ over $\mathcal{T}^+$ in a variational Bayesian neural architecture via minimizing Kullback-Leibler divergence~\cite{kullback1951information}.

To transfer dense vector representation of a subset of skills $v_\textbf{s}$, they construct a heterogeneous tripartite graph $\mathbb{G}=\langle \mathcal{N}, \mathcal{L}\rangle$ whose nodes are skills $n_{\mathcal{S}}$, successful teams $n_{\mathcal{T}^+}$ and experts $n_{\mathcal{E}}$, i.e., $\mathcal{N}=n_{\mathcal{S}}\cup n_{\mathcal{T}^+}\cup n_{\mathcal{E}}$, and links are undirected $\mathcal{L}=\{ n_{\mathcal{S}}\times n_{\mathcal{T}^+}\}\cup\{ n_{\mathcal{T}^+}\times n_{\mathcal{E}}\}$. From Figure~\ref{fig:hop-e2e} (right), each team $t_{\textbf{s},\textbf{e},y=1}\in\mathcal{T}^+$ is represented by a star subgraph with a team node as the central node and its required skills $\textbf{s}$ and expert members $\textbf{e}$ are the leaves, connected to the team node but not to each other. Then, a graph neural network $g$, parameterized by $\phi$, is trained such that $g_\phi: \mathcal{N}\rightarrow\mathbb{R}^d, g_\phi(n_{s_i})=v_{s_i}$. Finally, the dense vector of the required subset of skills $\textbf{s}$ for a team ${t}_{\textbf{s},\textbf{e},y}$ is obtained by summing the vectors of its constituent skill nodes, i.e., 
$v_\textbf{s} =\sum_{s_i \in \textbf{s}} v_{s_i}= \sum_{s_i \in \textbf{s}} g_\phi(n_{s_i})$.

\noindent\textbf{Limitations:} $g_\phi$ is estimated to capture structural and semantic information of skills only, completely disregarding the experts and their multi-hop collaborations within the context of historically successful teams. Moreover, it is \textit{prior} to and \textit{disjoint} from the main team recommendation task and is kept constant and oblivious to the multi-label classification loss when estimating $f_\theta$. Finally, the estimation of $f_\theta$, the primary predictor, relies on the mutually independent selection of teams from the training set.

\vspace{-0.5em}
\section{Proposed Approach}
\label{sec:approach}
We propose to estimate $f$ directly through the link prediction loss among \texttt{team-expert} links by $g_{\phi}$ on the expert collaboration graph, eliminating $\theta$ and the multi-label classifier. Following the standard message passing formulation and without loss of generality to modern graph neural networks, $g_{\phi}(n)$ denotes the vector representation of a node $n\in\mathcal{N}$ at layer (hop) $l$, and computed as:
\vspace{-0.5em}
\begin{equation}
\label{eq:gnn}
\!g_{\phi}(n)\!=\!\mathbf{h}_n^l
\!=\!
\phi_{update}\!\left(
\mathbf{h}_n^{(l-1)},
\!\!\!\!\sum_{i \in \mathcal{N}_n\subseteq\mathcal{N}}
\!\!\phi_{message}\!\left(
\mathbf{h}_n^{(l-1)},
\mathbf{h}_i^{(l-1)}
\right)\!\!
\right)
\end{equation}
where $\mathbf{h}_n^{l=0}$ denotes the randomly initialized vectors for the node $n$, $\mathcal{N}_n$ is the set of neighboring nodes of $n$, $\phi_{message}$ are learnable parameters for a message function to process the neighbors' vectors prior to aggregation, and ranges from a simple parameter-free identity function, as commonly used in variety of graph neural networks such as \texttt{gcn}~\cite{DBLP:conf/iclr/KipfW17} and \texttt{graphsage}~\cite{DBLP:conf/nips/HamiltonYL17}, to a more expressive multilayer perceptron with attention coefficients over neighboring nodes, as in \texttt{gat}~\cite{DBLP:conf/iclr/VelickovicCCRLB18}. Similarly, $\phi_{update}$ denotes the learnable parameters of an update function that computes the node vector $\mathbf{h}_n^l$ by combining its previous vector $\mathbf{h}_n^{(l-1)}$ with the aggregated vector from its neighborhood, and can be as simple as summation, or another multilayer perceptron, as in \texttt{gin}~\cite{DBLP:conf/iclr/XuHLJ19}. The complete set of parameters is given by $\phi=\phi_{update}\cup\phi_{message}$, which are learned by minimizing a link prediction loss over observed team–expert links of all successful teams, i.e., $\forall {t}_{\textbf{s},\textbf{e},y=1} \in \mathcal{T}^+, \forall e_j \in \textbf{e}$:
\begin{equation}
\label{eq:bxe}
\!\!\!\!\!\!\!\!\sum_{(n_{t_\textbf{s}},n_{e_j})\in \mathcal{L}}
\!\!\!\!\!\!\!\!\!-\texttt{log}\;\sigma\!\left(
g_{\phi}(n_{t_\textbf{s}})^\top g_{\phi}(n_{e_j})
\!\right) = \!\!\!\!\!\!\!\sum_{(n_{t_\textbf{s}},n_{e_j})\in \mathcal{L}}
\!\!\!\!\!\!\!\!\!\!-\texttt{log}\;\sigma\!\left(
{[\mathbf{h}_{n_{t_\textbf{s}}}^l]\!}^\top~[\mathbf{h}_{n_{e_j}}^l]
\!\right)
\end{equation}
where $\sigma$ is the sigmoid function, predicting the presence of links between a successful team and its expert members. While Eq.~\ref{eq:gnn} leverages \textit{all} types of links for message passing irrespective of their semantic roles, capturing heterogeneous relational and structural information, Eq.~\ref{eq:bxe} explicitly uses this information to supervise the primary task of team–expert link prediction for recommending successful teams.
During inference, as shown in Figure~\ref{fig:hop-e2e}~(right), for a team $t_{\textbf{s},*,y=1}$ with a subset of required skills \textbf{s} yet \textit{unknown} expert members, we use the learned $g_\phi$ to predict links between the team node $n_t$ and all candidate expert nodes $n_\mathcal{E}$. Specifically, for each expert $e \in \mathcal{E}$, we compute:
\begin{equation}
\label{eq:gnn-infer}
\!\!\!f_{\theta}(\textbf{s})\simeq f(\textbf{s} : \mathbb{G}, g_{\phi}) = \sigma\left(g_{\phi}(n_{t_\textbf{s}})\!^\top g_{\phi}(n_{e})\!\right)= \sigma\!\left(
{[\mathbf{h}_{n_{t_\textbf{s}}}^l]\!}^\top~[\mathbf{h}_{n_{e}}^l]
\!\right)
\end{equation}
and select the subset of experts with the top-$k$ predicted probabilities as the recommended team of size $k$.

Our proposed end-to-end formulation directly taps into multi-hop relational and structural information encoded in the graph for complex intra-team and cross-team interactions and dependencies among experts and their associated skills, as shown in Figure~\ref{fig:hop-e2e}~(left); a capacity in which neural classifiers are inherently limited. Moreover, it avoids two disjoint learning stages with separate parameter sets, i.e., the pretraining of $g_\phi$ and the subsequent fine-tuning for $f_\theta$. Herein, $\phi$ is the sole learnable set of parameters.


\section{Experiments}\label{sec:experiments}
In this section, we seek to answer our main research question:

\noindent\textbf{RQ1}: \textit{Does the end-to-end approach outperform the transfer-based approach for team recommendation?} 
If affirmative, we further ask:

\noindent\textbf{RQ2}: \textit{Is \textbf{RQ1}'s answer consistent across datasets from varied domains with distinct distributions of teams over skills?} 

\noindent\textbf{RQ3}: \textit{Is \textbf{RQ1}'s answer consistent across different graph structures?} 

\noindent\textbf{RQ4}: \textit{Which graph neural network performs the best (worst) in the end-to-end vs. transfer-based approaches for team recommendation?} 


\vspace{-0.5em}
\subsection{Datasets}
\begin{figure}[t]
\vspace{-1.5em}
\centering
\includegraphics[scale=0.34]{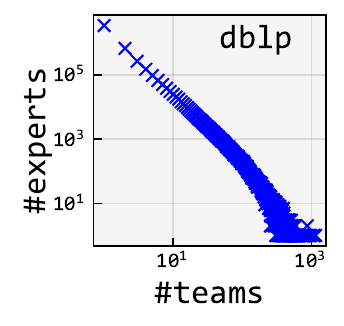} 
\includegraphics[scale=0.34]{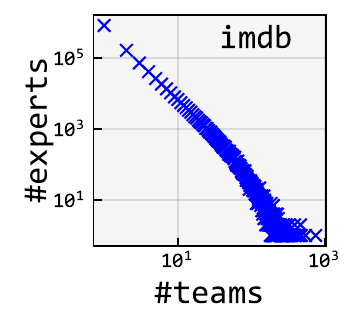}
\rulesep
\includegraphics[scale=0.34]{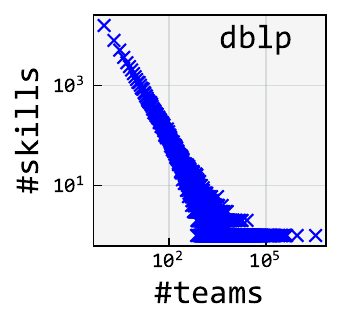}
\includegraphics[scale=0.34]{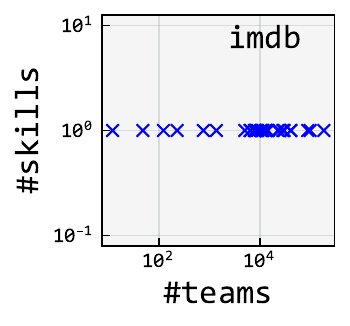}
\vspace{-1.5em}
\caption{Distribution of teams over experts and skills.}
\label{fig:distsbefore}
\vspace{-2em}
\end{figure}

We used two benchmark datasets as in prior works~\cite{DBLP:journals/tois/RadFBKSS24,DBLP:conf/cikm/DashtiSF22,DBLP:conf/cikm/RadFKSB20,DBLP:conf/icde/KouSSLN0M20,DBLP:journals/tkde/KargarGSSZ22}: 

\noindent\textbf{\texttt{dblp}}, a collection of computer science publications~\cite{dblp}, where a \textit{published} paper is a successful team, the authors are the expert members, and the keywords are the required skills, and 
\noindent\textbf{\texttt{imdb}}, a collection of movies~\cite{imdb}, where a \textit{produced} movie is considered as a successful team, the cast and crew are the experts, and the movie's genres and subgenres are the required skills. In contrast to movie recommenders or movie review analysis, herein, the objective is to form a team for a successful movie production. Teams with fewer than \texttt{3} expert members were removed while ensuring no alteration to the statistical distributions, as detailed in Table~\ref{tab:data_stats}. From Figure~\ref{fig:distsbefore} (left), both datasets exhibit long-tailed distributions of teams over experts, i.e., few dominant experts have contributed to many teams, while the majority have scarcely participated. Regarding skills, from Figure ~\ref{fig:distsbefore} (right), while \texttt{dblp} follows a similar long-tail distribution of teams over skills, \texttt{imdb} demonstrates a more uniform distribution across a limited set of skills (genres), which are consistently used in many movies.
\begin{table}[t]
\centering
\Large
\caption{Statistics of the raw and preprocessed datasets.}
\vspace{-1em}
\resizebox{0.6\columnwidth}{!}{\begin{tabular}{@{}lcc|cc@{}}
\cline{2-5}
& \multicolumn{2}{c}{\texttt{dblp}} & \multicolumn{2}{c}{\texttt{imdb}} \\\hline 

teams $\mathcal{T}$ & \multicolumn{2}{c}{publications} & \multicolumn{2}{c}{movies}\\

experts $\mathcal{E}$ & \multicolumn{2}{c}{authors} & \multicolumn{2}{c}{cast \& crew}\\

skills $\mathcal{S}$ & \multicolumn{2}{c}{keywords} & \multicolumn{2}{c}{(sub) genres} \\

success $y$ & \multicolumn{2}{c}{published} & \multicolumn{2}{c}{produced} \\\hline

statistics & raw & filtered & raw & filtered \\ \hline
\multicolumn{1}{@{}l}{|$\mathcal{T}$|} & 4,877,383 & 99,375 & 507,034 & 32,059 \\

\multicolumn{1}{@{}l}{|$\mathcal{E}$|} & 5,022,955 & 14,214 & 876,981 & 2,011 \\

\multicolumn{1}{@{}l}{|$\mathcal{S}$|} & \textbf{89,504} & \textbf{29,661} & \textbf{28} & \textbf{23} \\






\hline
\vspace{-1em}
\end{tabular}}

\label{tab:data_stats}
\end{table}
\vspace{-0.5em}
\subsection{Baselines}
We compared the end-to-end approach (\texttt{e2e-*}) with the following transfer-based (\texttt{t-*}) baselines:

\noindent\textbf{\texttt{t-bnn.m2v}}~\cite{DBLP:conf/sigir/RadBKSS21}: It is the pioneer to transfer embeddings of skills by \texttt{metapath2vec}~\cite{DBLP:conf/kdd/DongCS17} for a variational Bayesian neural classifier.
    
\noindent\texttt{\textbf{t-bnn.lant}}~\cite{DBLP:conf/ijcnn/KawKS23}: It employs deep graph infomax~\cite{DBLP:conf/iclr/VelickovicFHLBH19} for more effective embeddings of skills in fewer training epochs.

To ensure a comprehensive comparison between the end-to-end and transfer-based approaches, we also included strong graph neural networks for a heterogeneous graph\footnote{\href{https://pytorch-geometric.readthedocs.io/en/stable/generated/torch_geometric.nn.conv.HeteroConv.html}{\texttt{torch\_geometric.nn.conv.HeteroConv.html}}}:

\noindent\textbf{\texttt{graphsage (gs)}}~\cite{DBLP:conf/nips/HamiltonYL17}: It aggregates neighbours' vectors by a fixed function, herein, \texttt{mean}, followed by concatenation with the node's current vector.

\noindent\textbf{\texttt{gin}}~\cite{DBLP:conf/iclr/XuHLJ19}: It learns node vectors via \textit{learnable} aggregation and combination functions, which, theoretically, show the best node embeddings by capturing graph structure based on the Weisfeiler-Lehman graph isomorphism test~\cite{douglas2011weisfeiler}. 

\noindent\textbf{\texttt{gine}}~\cite{DBLP:conf/iclr/HuLGZLPL20}: It extends \texttt{gin} to include global graph features based on a subgraph of neighbors located between $k_1$-hop and $k_2$-hop of a node where $k_1<k_2$, and is designed to utilize link features.

\noindent\textbf{\texttt{gat}}~\cite{DBLP:conf/iclr/VelickovicCCRLB18}: It aggregates neighbors with \textit{learnable} multi-head attention scores based on the importance of neighbors when aggregating and combining vectors during message passing.

\noindent\textbf{\texttt{gatv2}}~\cite{DBLP:conf/iclr/Brody0Y22}: As opposed to \texttt{gat}, which computes pairwise attention between a node and its neighbors using shared parameters, it uses a different set of parameters for a node and each of its neighbors that are updated through layers (hops).

\noindent\textbf{\texttt{han}}~\cite{DBLP:conf/www/WangJSWYCY19}: It is a hierarchical \texttt{gat} that captures the importance of different node types and their predefined meta-paths to neighbours. 

\begin{figure}[t]
\centering


\includegraphics[scale=0.4]{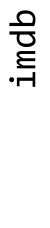} 
\hspace{-0.7em}
\includegraphics[scale=0.38]{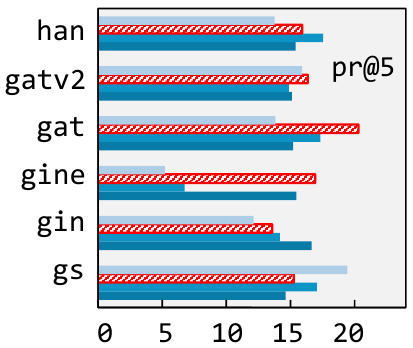} 
\includegraphics[scale=0.38]{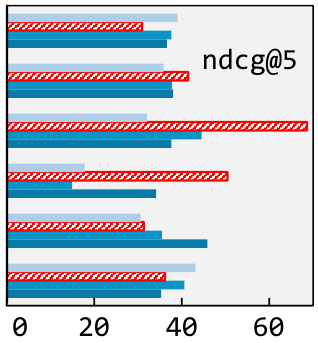}
\includegraphics[scale=0.38]{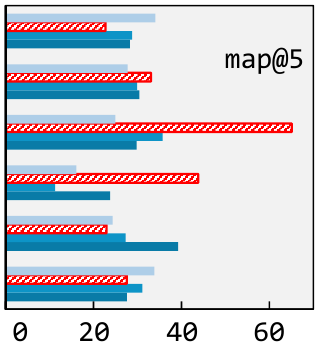}
\includegraphics[scale=0.32]{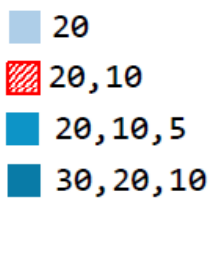}
\vspace{-1.em}
\caption{Impact of neighborhood sampling.}
\label{fig:sampling}
\vspace{-2.em}
\end{figure}

All graph neural networks included convolutional layers of size \textbf{\texttt{64}} and \texttt{relu} activation, after a search over different sizes of \{\texttt{16}, \texttt{32}, \texttt{64}, \texttt{128}, \texttt{256}\} across \texttt{dblp} and \texttt{imdb} datasets. For message passing, all neighbors were considered with a negative sampling at a ratio of \texttt{5:1}. We employed \texttt{1}-hop sampling with \texttt{20} nodes, \texttt{2}-hop sampling with [\texttt{20,10}], i.e., \texttt{20} and \texttt{10} nodes in the first and second hops, and \texttt{3}-hop sampling with [\texttt{20,10,5}] and [\texttt{30,20,10}]. From Figure~\ref{fig:sampling}, we observed generally favorable results with \textbf{\texttt{2}}-hop sampling and \textbf{[\texttt{20,10}]}, though not uniformly across all models. 

For the transfer-based, we cross-compare variational Bayesian (\texttt{t-bnn-*}) and non-variational (\texttt{t-fnn-*}) neural classifiers, following their settings in~\cite{DBLP:conf/sigir/RadBKSS21,DBLP:journals/tois/RadFBKSS24}. For the end-to-end approach, we adopt the same training hyperparameters (e.g., learning rate, number of epochs, and batch size) used in the transfer-based baselines for graph neural networks to ensure a fair comparison by controlling over- or under-tuning either approach relative to the other. 

\begin{table*}[t!]
\centering
\caption{Efficacy of end-to-end vs. transfer-based approaches in \texttt{dblp} and \texttt{imdb} using \texttt{skill-team-expert} graph structure.}
\label{tab:dblp_imdb_ste}
\vspace{-1.5em}
\resizebox{0.8\textwidth}{!}{%
\renewcommand{\arraystretch}{0.9}

\begin{tabular}{@{}l@{\hskip0.1cm}l@{\hskip0.15cm}c@{\hskip0.14cm}cc@{\hskip0.14cm}cc@{\hskip0.14cm}cc@{\hskip0.14cm}c|c@{\hskip0.14cm}cc@{\hskip0.14cm}cc@{\hskip0.14cm}cc@{\hskip0.14cm}c@{}}

\cline{3-18}
\multicolumn{2}{l}{} &  \multicolumn{8}{c}{\texttt{dblp}} & \multicolumn{8}{c}{\texttt{imdb}} \\ \cline{3-18}

\multicolumn{2}{l}{} & \multicolumn{2}{c}{\texttt{\%pr}} & \multicolumn{2}{c}{\texttt{\%rec}} & \multicolumn{2}{c}{\texttt{\%ndcg}} & \multicolumn{2}{c}{\texttt{\%map}} & \multicolumn{2}{c}{\texttt{\%pr}} & \multicolumn{2}{c}{\texttt{\%rec}} & \multicolumn{2}{c}{\texttt{\%ndcg}} & \multicolumn{2}{c}{\texttt{\%map}} \\ \cline{3-18}

\multicolumn{2}{l}{\;\;\;$k$} & \texttt{@5} &  \texttt{@10} &  \texttt{@5} &  \texttt{@10} &  \texttt{@5} &  \texttt{@10}&  \texttt{@5}  &  \texttt{@10}&  \texttt{@5}  &  \texttt{@10} &  \texttt{@5} &  \texttt{@10} &  \texttt{@5} &  \texttt{@10} &  \texttt{@5} &  \texttt{@10} 

\\ \hline

\multirowvertlinecustom{6}{1.2}{-0.4}{0.3}&\texttt{gs}& \cellcolor[HTML]{DEEBF9}14.37          & \cellcolor[HTML]{D9E8F7}09.24          & \cellcolor[HTML]{D6E7F7}54.00          & \cellcolor[HTML]{D0E3F6}67.79          & \cellcolor[HTML]{DFECF9}33.21          & \cellcolor[HTML]{DCEAF8}38.39          & \cellcolor[HTML]{E3EEFA}25.18          & \cellcolor[HTML]{E1EDF9}28.10          & \cellcolor[HTML]{D6E7F7}15.39          & \cellcolor[HTML]{DAE9F8}09.16          & \cellcolor[HTML]{CCE1F5}49.73          & \cellcolor[HTML]{D1E3F6}57.44          & \cellcolor[HTML]{D2E4F6}33.70          & \cellcolor[HTML]{D4E5F7}36.78          & \cellcolor[HTML]{D3E5F6}27.00          & \cellcolor[HTML]{D5E6F7}29.03          \\

&\texttt{gin}& \cellcolor[HTML]{DEEBF9}{\ul ~14.47}    & \cellcolor[HTML]{D9E8F8}09.07          & \cellcolor[HTML]{D6E6F7}{\ul ~54.68}    & \cellcolor[HTML]{D0E3F6}66.95          & \cellcolor[HTML]{E0EDF9}31.50          & \cellcolor[HTML]{DEEBF9}36.12          & \cellcolor[HTML]{E5EFFA}22.74          & \cellcolor[HTML]{E3EEFA}25.36          & \cellcolor[HTML]{D9E8F8}14.08          & \cellcolor[HTML]{DBE9F8}08.91          & \cellcolor[HTML]{D0E3F6}45.34          & \cellcolor[HTML]{D1E4F6}56.49          & \cellcolor[HTML]{D8E8F7}28.49          & \cellcolor[HTML]{D8E8F7}32.79          & \cellcolor[HTML]{DAE9F8}21.76          & \cellcolor[HTML]{DAE9F8}24.30          \\

&\texttt{gine}& \cellcolor[HTML]{E9F2FB}07.66          & \cellcolor[HTML]{DEEBF9}07.65          & \cellcolor[HTML]{E9F2FB}21.56          & \cellcolor[HTML]{DDEBF8}44.17          & \cellcolor[HTML]{EAF3FB}16.50          & \cellcolor[HTML]{E5F0FA}24.74          & \cellcolor[HTML]{ECF4FC}12.67          & \cellcolor[HTML]{EAF2FB}16.18          & \cellcolor[HTML]{CEE2F5}{\ul ~19.33}    & \cellcolor[HTML]{D0E3F6}{\ul ~12.21}    & \cellcolor[HTML]{C9DFF4}{\ul ~53.14}    & \cellcolor[HTML]{CBE0F5}{\ul ~66.25}    & \cellcolor[HTML]{CDE1F5}{\ul ~38.80}    & \cellcolor[HTML]{CDE1F5}{\ul ~43.86}    & \cellcolor[HTML]{CEE1F5}{\ul ~31.13}    & \cellcolor[HTML]{CFE2F5}{\ul ~34.07}    \\

&\texttt{gat}& \cellcolor[HTML]{DEEBF9}14.33          & \cellcolor[HTML]{D8E8F7}09.46          & \cellcolor[HTML]{D7E7F7}53.81          & \cellcolor[HTML]{CFE2F6}68.86          & \cellcolor[HTML]{DCEAF8}{\ul ~38.11}    & \cellcolor[HTML]{D8E8F7}43.78          & \cellcolor[HTML]{DEEBF9}{\ul ~31.44}    & \cellcolor[HTML]{DCEAF8}34.65          & \cellcolor[HTML]{D8E8F7}14.66          & \cellcolor[HTML]{DAE9F8}09.17          & \cellcolor[HTML]{CEE2F5}47.44          & \cellcolor[HTML]{D1E3F6}57.47          & \cellcolor[HTML]{D3E4F6}33.45          & \cellcolor[HTML]{D4E5F6}37.40          & \cellcolor[HTML]{D3E5F6}27.17          & \cellcolor[HTML]{D4E5F7}29.68          \\

&\texttt{gatv2}& \cellcolor[HTML]{DFECF9}14.03          & \cellcolor[HTML]{D8E8F7}{\ul ~09.48}    & \cellcolor[HTML]{D7E7F7}52.73          & \cellcolor[HTML]{CFE2F6}{\ul ~69.01}    & \cellcolor[HTML]{DCEAF8}37.87          & \cellcolor[HTML]{D8E8F7}{\ul ~43.98}    & \cellcolor[HTML]{DEEBF9}31.43          & \cellcolor[HTML]{DCEAF8}{\ul ~34.86}    & \cellcolor[HTML]{D8E8F7}14.49          & \cellcolor[HTML]{DAE9F8}09.09          & \cellcolor[HTML]{CFE2F5}46.72          & \cellcolor[HTML]{D1E3F6}57.18          & \cellcolor[HTML]{D2E4F6}33.81          & \cellcolor[HTML]{D3E5F6}37.88          & \cellcolor[HTML]{D2E4F6}27.86          & \cellcolor[HTML]{D3E5F6}30.39          \\

\multirow{-6}{*}{\rotatebox[origin=c]{90}{\texttt{e2e}}} &
  \texttt{han} & \cellcolor[HTML]{C5DCF3}\textbf{28.97} & \cellcolor[HTML]{C5DCF3}\textbf{15.17} & \cellcolor[HTML]{C5DCF3}\textbf{83.19} & \cellcolor[HTML]{C5DCF3}\textbf{86.38} & \cellcolor[HTML]{C5DCF3}\textbf{70.42} & \cellcolor[HTML]{C5DCF3}\textbf{71.74} & \cellcolor[HTML]{C5DCF3}\textbf{64.28} & \cellcolor[HTML]{C5DCF3}\textbf{65.27} & \cellcolor[HTML]{C5DCF3}\textbf{23.37} & \cellcolor[HTML]{C5DCF3}\textbf{15.42} & \cellcolor[HTML]{C5DCF3}\textbf{57.82} & \cellcolor[HTML]{C5DCF3}\textbf{74.15} & \cellcolor[HTML]{C5DCF3}\textbf{45.58} & \cellcolor[HTML]{C5DCF3}\textbf{52.38} & \cellcolor[HTML]{C5DCF3}\textbf{37.42} & \cellcolor[HTML]{C5DCF3}\textbf{42.21} \\

  \hline\hline

\multirowvertlinecustom{8}{1.9}{-0.4}{0.3}


&\texttt{m2v}&00.82&00.70&\cellcolor[HTML]{F4F9FD}01.24&\cellcolor[HTML]{F4F9FD}02.13&\cellcolor[HTML]{F4F9FD}01.13&\cellcolor[HTML]{F4F9FD}01.55&00.66&00.82&00.84&00.77&00.95&\cellcolor[HTML]{F4F9FD}01.76&00.95&\cellcolor[HTML]{F4F9FD}01.33&00.46&00.57\\

&\texttt{lant}&00.90&00.83&\cellcolor[HTML]{F4F9FD}01.35&\cellcolor[HTML]{F4F9FD}02.48&\cellcolor[HTML]{F4F9FD}01.17&\cellcolor[HTML]{F4F9FD}01.69&00.67&00.84&00.74&00.68&00.81&\cellcolor[HTML]{F4F9FD}01.52&00.85&\cellcolor[HTML]{F4F9FD}01.18&00.40&00.50\\

&\texttt{gs}&00.87&00.74&\cellcolor[HTML]{F4F9FD}01.30&\cellcolor[HTML]{F4F9FD}02.24&\cellcolor[HTML]{F4F9FD}01.26&\cellcolor[HTML]{F4F9FD}01.63&00.60&00.76&00.83&00.73&00.91&\cellcolor[HTML]{F4F9FD}01.64&00.95&\cellcolor[HTML]{F4F9FD}01.27&00.44&00.55\\

&\texttt{gin}&00.95&00.78&\cellcolor[HTML]{F4F9FD}01.43&\cellcolor[HTML]{F4F9FD}02.35&\cellcolor[HTML]{F4F9FD}01.37&\cellcolor[HTML]{F4F9FD}01.57&00.59&00.76&00.76&00.68&00.84&\cellcolor[HTML]{F4F9FD}01.56&00.87&\cellcolor[HTML]{F4F9FD}01.19&00.40&00.51\\

&\texttt{gine}&\cellcolor[HTML]{F4F9FD}01.02&00.83&\cellcolor[HTML]{F4F9FD}01.52&\cellcolor[HTML]{F4F9FD}02.49&\cellcolor[HTML]{F4F9FD}01.35&\cellcolor[HTML]{F4F9FD}01.80&00.76&00.92&00.73&00.64&00.82&\cellcolor[HTML]{F4F9FD}01.45&00.85&\cellcolor[HTML]{F4F9FD}01.14&00.41&00.50\\

&\texttt{gat}&\cellcolor[HTML]{F4F9FD}01.19&00.92&\cellcolor[HTML]{F4F9FD}01.78&\cellcolor[HTML]{F4F9FD}02.75&\cellcolor[HTML]{F4F9FD}01.62&\cellcolor[HTML]{F4F9FD}02.08&00.92&\cellcolor[HTML]{F4F9FD}01.09&00.92&00.82&\cellcolor[HTML]{F4F9FD}01.00&\cellcolor[HTML]{F4F9FD}01.83&\cellcolor[HTML]{F4F9FD}01.06&\cellcolor[HTML]{F4F9FD}01.43&00.50&00.62\\

&\texttt{gatv2}&\cellcolor[HTML]{F4F9FD}01.01&00.83&\cellcolor[HTML]{F4F9FD}01.51&\cellcolor[HTML]{F4F9FD}02.51&\cellcolor[HTML]{F4F9FD}01.41&\cellcolor[HTML]{F4F9FD}01.88&00.80&00.96&00.87&00.80&00.96&\cellcolor[HTML]{F4F9FD}01.80&00.98&\cellcolor[HTML]{F4F9FD}01.37&00.47&00.59\\

\multirow{-8}{*}{\rotatebox[origin=c]{90}{\texttt{t-fnn}}}&\texttt{han}&00.93&00.82&\cellcolor[HTML]{F4F9FD}01.42&\cellcolor[HTML]{F4F9FD}02.49&\cellcolor[HTML]{F4F9FD}01.33&\cellcolor[HTML]{F4F9FD}01.82&00.80&00.97&00.77&00.72&00.86&\cellcolor[HTML]{F4F9FD}01.63&00.85&\cellcolor[HTML]{F4F9FD}01.21&00.39&00.50\\

\hline\hline
\multirowvertlinecustom{8}{1.9}{-0.4}{0.3}


&\texttt{m2v}&00.31&00.32&00.48&00.98&00.42&00.65&00.23&00.30&00.76&00.71&00.83&\cellcolor[HTML]{F5F9FD}01.58&00.85&\cellcolor[HTML]{F5F9FD}01.20&00.41&00.52\\

&\texttt{lant}&00.66&00.61&\cellcolor[HTML]{F5F9FD}01.01&\cellcolor[HTML]{F5F9FD}01.87&00.88&\cellcolor[HTML]{F5F9FD}01.28&00.49&00.61&\cellcolor[HTML]{F5F9FD}01.01&00.96&\cellcolor[HTML]{F5F9FD}01.14&\cellcolor[HTML]{F5F9FD}02.18&\cellcolor[HTML]{F5F9FD}01.18&\cellcolor[HTML]{F5F9FD}01.66&00.58&00.73\\

&\texttt{gs}&00.71&00.66&\cellcolor[HTML]{F4F9FD}01.08&\cellcolor[HTML]{F4F9FD}02.00&00.94&\cellcolor[HTML]{F4F9FD}01.36&00.52&00.65&00.99&00.90&\cellcolor[HTML]{F4F9FD}01.12&\cellcolor[HTML]{F4F9FD}02.03&\cellcolor[HTML]{F4F9FD}01.12&\cellcolor[HTML]{F4F9FD}01.55&00.54&00.67\\

&\texttt{gin}&00.60&00.52&00.91&\cellcolor[HTML]{F5F9FD}01.59&00.84&\cellcolor[HTML]{F5F9FD}01.15&00.49&00.60&00.90&00.83&\cellcolor[HTML]{F5F9FD}01.04&\cellcolor[HTML]{F5F9FD}01.92&\cellcolor[HTML]{F5F9FD}01.07&\cellcolor[HTML]{F5F9FD}01.47&00.52&00.65\\

&\texttt{gine}&00.75&00.63&\cellcolor[HTML]{F4F9FD}01.12&\cellcolor[HTML]{F4F9FD}01.90&\cellcolor[HTML]{F4F9FD}01.08&\cellcolor[HTML]{F4F9FD}01.44&00.66&00.79&\cellcolor[HTML]{F4F9FD}01.01&00.93&\cellcolor[HTML]{F4F9FD}01.15&\cellcolor[HTML]{F4F9FD}02.14&\cellcolor[HTML]{F4F9FD}01.16&\cellcolor[HTML]{F4F9FD}01.62&00.58&00.71\\

&\texttt{gat}&00.66&00.62&00.99&\cellcolor[HTML]{F5F9FD}01.86&00.87&\cellcolor[HTML]{F5F9FD}01.28&00.49&00.62&00.96&00.86&\cellcolor[HTML]{F5F9FD}01.07&\cellcolor[HTML]{F5F9FD}01.94&\cellcolor[HTML]{F5F9FD}01.10&\cellcolor[HTML]{F5F9FD}01.50&00.53&00.66\\

&\texttt{gatv2}& 00.64& 00.54& 00.96& \cellcolor[HTML]{F5F9FD}01.64& 00.87& \cellcolor[HTML]{F5F9FD}01.19& 00.49& 00.58& 00.93& 00.80& \cellcolor[HTML]{F5F9FD}01.05& \cellcolor[HTML]{F5F9FD}01.84& \cellcolor[HTML]{F5F9FD}01.05& \cellcolor[HTML]{F5F9FD}01.41& 00.51& 00.62\\

\multirow{-8}{*}{\rotatebox[origin=c]{90}{\texttt{t-bnn}}}&\texttt{han}& 00.80& 00.71& \cellcolor[HTML]{F4F9FD}01.20& \cellcolor[HTML]{F4F9FD}02.15& \cellcolor[HTML]{F4F9FD}01.09& \cellcolor[HTML]{F4F9FD}01.53& 00.61& 00.76& \cellcolor[HTML]{F4F9FD}01.06& 00.99& \cellcolor[HTML]{F4F9FD}01.19& \cellcolor[HTML]{F4F9FD}02.25& \cellcolor[HTML]{F4F9FD}01.21& \cellcolor[HTML]{F4F9FD}01.70& 00.59& 00.74\\
\hline

\end{tabular}%
}
\vspace{-1.5em}
\end{table*}
\vspace{-0.5em}
\subsection{Evaluation Strategy and Metrics}
We used the entire dataset to create the expert collaboration graph. For the test set, we selected \texttt{15}\% of the randomly shuffled teams and removed their \texttt{team-expert} links $(n_t, n_e)\in\mathcal{L}$ to mask the expert members as \textit{unseen} during training. We then performed \texttt{3}-fold cross-validation on the remaining \texttt{85}\% teams by further masking \texttt{team-expert} links for teams within each validation fold, resulting in one trained model per fold. For reproducibility and fair comparison, an identical benchmark with a fixed random seed is used across all methods. 
Given a team $t_{\textbf{s},\textbf{e},y=1}$ from the test set, a trained model infers the membership probability of all experts by \texttt{team-expert} link predictions. We selected the top-$k\!\in\!\{5,10\}$ experts as the recommended team and compared them against the ground-truth expert set $\textbf{e}$ and reported the average performance of models on all folds in terms of precision (\texttt{pr}), recall (\texttt{rec}), normalized discounted cumulative gain (\texttt{ndcg}), and mean average precision (\texttt{map}). 
\begin{table}[]
\centering
\caption{Efficacy of approaches using \texttt{location} in \texttt{dblp}.}
\label{tab:dblp_imdb_loc}
\vspace{-1.5em}
 
\resizebox{0.84\columnwidth}{!}{%
\begin{tabular}{@{}l@{\hskip0.1cm}l@{\hskip0.01cm}c@{\hskip0.16cm}cc@{\hskip0.15cm}cc@{\hskip0.15cm}cc@{\hskip0.15cm}c@{}}\cline{3-10}
  
\multicolumn{2}{l}{} & \multicolumn{2}{c}{\texttt{\%pr}} & \multicolumn{2}{c}{\texttt{\%rec}} & \multicolumn{2}{c}{\texttt{\%ndcg}} & \multicolumn{2}{c}{\texttt{\%map}} \\ \cline{3-10}

 & $k$& \texttt{@5} & \texttt{@10} &  \texttt{@5} &  \texttt{@10} &  \texttt{@5} &  \texttt{@10} &  \texttt{@5} &  \texttt{@10} \\
 
 \hline\hline

 \multirowvertlinecustom{8}{3}{1.2}{0.3}&\texttt{gs} & \cellcolor[HTML]{C8DEF4}{\ul ~14.63}    & \cellcolor[HTML]{C7DDF4}09.15          & \cellcolor[HTML]{C8DEF4}{\ul ~52.91}    & \cellcolor[HTML]{C7DDF4}64.97          & \cellcolor[HTML]{CEE1F5}35.37          & \cellcolor[HTML]{CCE0F5}39.90          & \cellcolor[HTML]{D1E3F6}28.27          & \cellcolor[HTML]{CFE2F6}30.88          \\

 &\texttt{gin} & \cellcolor[HTML]{D1E4F6}11.80          & \cellcolor[HTML]{C9DEF4}08.78          & \cellcolor[HTML]{D1E4F6}42.44          & \cellcolor[HTML]{C8DEF4}62.46          & \cellcolor[HTML]{D8E8F7}26.28          & \cellcolor[HTML]{D3E4F6}33.65          & \cellcolor[HTML]{DCEAF8}19.51          & \cellcolor[HTML]{D9E8F7}23.54          \\


 &\texttt{gat} & \cellcolor[HTML]{CBE0F5}13.69          & \cellcolor[HTML]{C5DCF3}\textbf{09.39} & \cellcolor[HTML]{CBE0F5}50.24          & \cellcolor[HTML]{C5DCF3}\textbf{66.50} & \cellcolor[HTML]{CEE1F5}{\ul ~35.57}    & \cellcolor[HTML]{CADFF4}{\ul ~41.74}    & \cellcolor[HTML]{CFE3F6}{\ul ~29.10}    & \cellcolor[HTML]{CDE1F5}{\ul ~32.62}    \\

 &\texttt{gatv2~~~~~~} & \cellcolor[HTML]{C5DCF3}\textbf{15.46} & \cellcolor[HTML]{C6DDF4}{\ul ~09.22}    & \cellcolor[HTML]{C5DCF3}\textbf{56.15} & \cellcolor[HTML]{C6DDF4}{\ul ~65.38}    & \cellcolor[HTML]{C5DCF3}\textbf{42.58} & \cellcolor[HTML]{C5DCF3}\textbf{46.13} & \cellcolor[HTML]{C5DCF3}\textbf{36.62} & \cellcolor[HTML]{C5DCF3}\textbf{38.75} \\

\multirow{-5}{*}{\rotatebox[origin=c]{90}{\texttt{e2e}}~~~~~~} & \texttt{han} & \cellcolor[HTML]{E1EDF9}06.77          & \cellcolor[HTML]{CDE1F5}07.93          & \cellcolor[HTML]{EAF3FB}13.90          & \cellcolor[HTML]{E1EDF9}29.40          & \cellcolor[HTML]{E6F0FA}13.85          & \cellcolor[HTML]{E1EDF9}20.50          & \cellcolor[HTML]{E8F2FB}10.15          & \cellcolor[HTML]{E5F0FA}13.39          \\

\hline\hline





 \multirowvertlinecustom{8}{3}{1.2}{0.3}&\texttt{gs}&00.54&00.54&00.82&01.65&00.71&\cellcolor[HTML]{F4F9FD}01.10&00.38&00.50\\

&\texttt{gin}&00.94&00.80&01.43&02.44&\cellcolor[HTML]{F4F9FD}01.28&\cellcolor[HTML]{F4F9FD}01.75&00.73&00.88\\


&\texttt{gat}&00.77&00.70&01.18&02.12&\cellcolor[HTML]{F4F9FD}01.04&\cellcolor[HTML]{F4F9FD}01.47&00.57&00.71\\

&\texttt{gatv2}&00.96&00.79&\cellcolor[HTML]{F4F9FD}01.45&\cellcolor[HTML]{F4F9FD}02.40&\cellcolor[HTML]{F4F9FD}01.25&\cellcolor[HTML]{F4F9FD}01.69&00.69&00.83\\

\multirow{-5}{*}{\rotatebox[origin=c]{90}{\texttt{t-fnn}}}&\texttt{han}&00.92&00.79&01.38&\cellcolor[HTML]{F4F9FD}02.36&\cellcolor[HTML]{F4F9FD}01.26&\cellcolor[HTML]{F4F9FD}01.72&00.72&00.87\\

\hline\hline





\multirowvertlinecustom{8}{3}{1.2}{0.3}&\texttt{gs}&00.85&00.73&\cellcolor[HTML]{F4F9FD}01.28&\cellcolor[HTML]{F4F9FD}02.20&\cellcolor[HTML]{F4F9FD}01.14&\cellcolor[HTML]{F4F9FD}01.56&00.64&00.78\\

&\texttt{gin}&00.46&00.42&00.70&\cellcolor[HTML]{F4F9FD}01.27&00.62&00.89&00.34&00.42\\


&\texttt{gat}&00.59&00.53&00.90&\cellcolor[HTML]{F4F9FD}01.61&00.81&\cellcolor[HTML]{F4F9FD}01.14&00.46&00.57\\

&\texttt{gatv2}&00.52&00.52&00.77&\cellcolor[HTML]{F4F9FD}01.56&00.67&\cellcolor[HTML]{F4F9FD}01.04&00.37&00.48\\

\multirow{-5}{*}{\rotatebox[origin=c]{90}{\texttt{t-bnn}}}&\texttt{han}&00.62&00.58&00.94&\cellcolor[HTML]{F4F9FD}01.76&00.81&\cellcolor[HTML]{F4F9FD}01.19&00.44&00.56\\
\hline
\end{tabular}%
}
\vspace{-2.5em}
\end{table}

\vspace{-0.5em}
\section{Results}\label{sec:results}

\noindent\textbf{RQ1: End-to-end vs. transfer-base approach.} From Table~\ref{tab:dblp_imdb_ste}, we observe that the end-to-end approach (\texttt{e2e-*}) consistently and substantially outperforms \textit{all} transfer-based methods using either \texttt{t-fnn-*} or \texttt{t-bnn-*} classifiers across all graph neural networks, datasets, and metrics. The superiority of the end-to-end approach lies in leveraging both multi-hop relational and structural information in the expert collaboration graph and supervised information about the optimal subset of experts within successful teams.

\noindent\textbf{RQ2: Cross-domain performance consistency.} From Table~\ref{tab:dblp_imdb_ste}, we can observe that the performance improvement by graph neural networks is consistent for both datasets in our end-to-end approach. However, in the transfer-based approach, the performance varies in variational and non-variational models with \textit{no} consistent trend based on their underlying graph neural network for different datasets. Yet, attentive models, \texttt{t-*-han}, \texttt{t-*-gat}, or \texttt{t-*-gatv2}, generally stand out as the top-performers.

\noindent\textbf{RQ3: Cross-graph structural performance consistency.} To study the impact of graph structure, we considered \textit{i}) \texttt{skill-team-} \texttt{expert-\textbf{location}} where a team is further connected to its location, and \textit{ii}) \texttt{skill-expert} bipartite graph, consisting of links between skills and expert members of a team but with \textit{no} node for the team. Following the literature~\cite{DBLP:conf/sigir/RadBKSS21}, we set the location of a team in \texttt{dblp} as the publication venue. In \texttt{imdb}, however, the available location information is the producing country of a movie, which is US in almost all movies, and was therefore omitted from our experiments. From Tables~\ref{tab:dblp_imdb_ste} and~\ref{tab:dblp_imdb_loc}, we observe that the performance of \texttt{e2e-*} models on the expert collaboration graph with location and lack thereof is closely aligned, with marginal differences, except for \texttt{han}, which is no longer the best due to its meta-path-based attention mechanism that overlooked the location links. In contrast, for transfer-based models, we observe little to \textit{no} improvement, which is expected as the vector representations of skills are \textit{in}sufficient to convey such additional information to neural classifiers. In the \texttt{skill-expert} bipartite graph, from Table~\ref{tab:skill-expert} including top-performing baselines, we observe that \textit{neither} end-to-end \textit{nor} transfer-based models perform on par with their performance in other graph structures due to the \textit{highly dense} bipartite graph structure, wherein all skills and experts are connected to each other without the contextual separation by team nodes, which highlights the importance of graph structure in effective message passing in graph neural networks.
\begin{table}[]
\centering
\caption{Inefficacy of \texttt{skill-expert} graph structure.}
\label{tab:skill-expert}
\vspace{-1.5em}
 
\resizebox{0.85\columnwidth}{!}{%
\begin{tabular}{@{}l@{\hskip0.1cm}l@{\hskip0.01cm}c@{\hskip0.15cm}c@{\hskip0.01cm}c@{\hskip0.15cm}c|c@{\hskip0.2cm}c@{\hskip0.015cm}c@{\hskip0.01cm}c@{}c}\cline{3-10}
\multicolumn{2}{l}{} & \multicolumn{4}{c}{\texttt{dblp}} &  \multicolumn{4}{c}{\texttt{imdb}}\\
\cline{3-10}
  
\multicolumn{2}{l}{$k=10$} & \multicolumn{1}{c}{\texttt{\%pr}} & \multicolumn{1}{c}{\texttt{\%rec}} & \multicolumn{1}{c}{\texttt{\%ndcg}} & \multicolumn{1}{c}{\texttt{\%map}}& \multicolumn{1}{c}{\texttt{\%pr}} & \multicolumn{1}{c}{\texttt{\%rec}} & \multicolumn{1}{c}{\texttt{\%ndcg}} & \multicolumn{1}{c}{\texttt{\%map}} \\

 \hline\hline

 \multirowvertlinecustom{6}{1.8}{0.9}{0.3}&\texttt{gs} & 00.04 & 00.15 & 00.08 & 00.04 & 00.13 & 00.42 & 00.28 & 00.14 \\


&\texttt{gat} & 00.03 & 00.13 & 00.07 & 00.03 & 00.21 & 00.64 & 00.39 & 00.16 \\

\multirow{-3}{*}{\rotatebox[origin=c]{90}{\texttt{e2e}}~~~~~~} &\texttt{han} & 00.05 & 00.18 & 00.10 & 00.04 & 00.15 & 00.49 & 00.32 & 00.15  \\

\hline\hline

 \multirowvertlinecustom{10}{3.5}{2.2}{0.3}&\texttt{m2v}&00.80 & \cellcolor[HTML]{F3F8FD}02.44 & \cellcolor[HTML]{F3F8FD}01.83 & 00.99 & 00.73 & \cellcolor[HTML]{F4F8FD}01.64 & \cellcolor[HTML]{F3F8FD}01.26 & 00.54 \\

&\texttt{gs}&00.83 & \cellcolor[HTML]{F3F8FD}02.51 & \cellcolor[HTML]{F3F8FD}01.46 & 00.71 & 00.75 & \cellcolor[HTML]{F3F8FD}01.70 & \cellcolor[HTML]{F3F8FD}01.31 & 00.56 \\


&\texttt{gat}&00.85 & \cellcolor[HTML]{F2F7FC}02.57 & \cellcolor[HTML]{F4F8FD}01.92 & 00.98 & 00.65 & \cellcolor[HTML]{F4F8FD}01.46 & \cellcolor[HTML]{F4F8FD}01.15 & 00.50 \\

\multirow{-4}{*}{\rotatebox[origin=c]{90}{\texttt{t-fnn}}}&\texttt{han} &00.84 & \cellcolor[HTML]{F2F7FC}02.54 & \cellcolor[HTML]{F2F7FC}01.81 & 00.95 & 00.73 & \cellcolor[HTML]{F2F7FC}01.61 & \cellcolor[HTML]{F2F7FC}01.25 & 00.53\\

\hline\hline

\multirowvertlinecustom{10}{3.5}{2.2}{0.3}&\texttt{m2v}&00.39 & \cellcolor[HTML]{F4F9FD}01.20 & 00.83 & 00.40 & 00.89 & \cellcolor[HTML]{F3F8FD}02.04 & \cellcolor[HTML]{F3F8FD}01.56 & 00.69 \\

&\texttt{gs}&00.58 & \cellcolor[HTML]{F3F8FD}01.75 & \cellcolor[HTML]{F4F8FD}01.22 & 00.59 & 00.96 & \cellcolor[HTML]{F3F8FD}02.18 & \cellcolor[HTML]{F3F8FD}01.61 & 00.68 \\

&\texttt{gat}&00.53 & \cellcolor[HTML]{F4F8FD}01.60 & \cellcolor[HTML]{F4F8FD}01.06 & 00.49 & \cellcolor[HTML]{F2F7FC}01.01 & \cellcolor[HTML]{F2F7FC}02.31 & \cellcolor[HTML]{F2F7FC}01.74 & 00.76\\

\multirow{-4}{*}{\rotatebox[origin=c]{90}{\texttt{t-bnn}}} &\texttt{han} & 00.51 & \cellcolor[HTML]{F4F8FD}01.54 & \cellcolor[HTML]{F4F8FD}01.05 & 00.51 & 00.94 & \cellcolor[HTML]{F4F8FD}02.15 & \cellcolor[HTML]{F4F8FD}01.64 & 00.72\\
\hline

\end{tabular}%
}
\vspace{-2.em}
\end{table}

\noindent\textbf{RQ4: Performance leaderboard for graph neural networks.} From Table~\ref{tab:dblp_imdb_ste} and for the end-to-end approach, it is evident that \texttt{han} outperforms \textit{all} methods across datasets and metrics, with \texttt{gat} and \texttt{gatv2} being the closest runners-up in \texttt{dblp}. The consistent high performance of \texttt{han}, \texttt{gat}, and \texttt{gatv2} highlights the dominance of attention-based graph neural networks. In contrast, despite the \texttt{gin} model's theoretically proven efficacy in capturing graph structure~\cite{DBLP:conf/iclr/XuHLJ19}, its empirical performance typically lags behind the attentive models for its sensitivity to hyperparameters. With respect to the transfer-based approach, not unexpectedly, we observe a consistent efficacy of graph neural networks compared to naive embeddings of skills. Attentive models (\texttt{gat}, \texttt{gatv2}) generally perform the best, similarly when employed in the end-to-end approach. However, the performance ranking of graph neural networks is \textit{in}consistent across neural classifiers; it is more stable for non-variational models (\texttt{t-fnn-*}) than for variational ones (\texttt{t-bnn-*}) due to the greater sensitivity of variational architectures.

\vspace{-0.5em}
\section{Concluding Remarks and Future Work}\label{sec:conclusion}
In this paper, we proposed an end-to-end graph neural network to improve the efficacy of team recommendations. Our experiments on two large-scale datasets with distinct distributions of teams over skills show state-of-the-art performance, regardless of the underlying graph neural network. For future work, we study cold-start scenarios for new experts and emerging skills with limited historical data, as well as team \textit{refinement}, where experts in an existing team are replaced to maintain or enhance performance.


\section{GenAI Usage Disclosure}
We used GenAI tools during manuscript preparation to rephrase text, improve sentence structure, and enhance grammar and fluency, ensuring clarity and coherence without altering the original meaning or technical content. They were also used to help develop scripts to run the benchmarks in parallel, understanding the characteristics of the graph neural networks, and to establish a solid, reproducible pipeline. Additionally, GenAI tools were used to assist in collecting bulk experimental results from raw output files.
\bibliographystyle{ACM-Reference-Format}
\bibliography{biball}
\end{document}